\documentclass[runningheads]{llncs}
\usepackage{siunitx}

\mathchardef\mhyphen="2D
\newcolumntype{C}{@{\extracolsep{6pt}}c@{\extracolsep{3pt}}}%
\newcolumntype{L}{@{\extracolsep{6pt}}l@{\extracolsep{3pt}}}%

\makeatletter
\DeclareRobustCommand\onedot{\futurelet\@let@token\@onedot}

\makeatother
\usepackage{graphicx}
\usepackage{makecell}
\usepackage{arydshln}
\usepackage{booktabs}
\usepackage{multirow}
\usepackage[bottom]{footmisc}
\usepackage{xcolor}
\usepackage{amsfonts,amssymb}
\usepackage{svg}
\newcommand{\M}{OA-BreaCR} 

\begin{document}
%
\title{
Ordinal Learning: Longitudinal Attention Alignment Model for Predicting Time to Future Breast Cancer Events from Mammograms
}
%
\titlerunning{Estimating Time to Future Breast Cancer Using Longitudinal Mammograms}
%
\author{
Xin Wang\inst{1,2} \and  
Tao Tan \inst{3,1 *} \and  
Yuan Gao\inst{1,2} \and  
Eric Marcus\inst{1} \and  
Luyi Han\inst{1,4} \and  
Antonio Portaluri\inst{1,5} \and  
Tianyu Zhang\inst{1,2,4} \and  
Chunyao Lu\inst{1,4} \and  
Xinglong Liang\inst{1,4} \and  
Regina Beets-Tan\inst{1,2} \and  
Jonas Teuwen\inst{1,4}  
Ritse Mann\inst{1,4}  
}

\authorrunning{Wang et al.}
%
\institute{
Department of Radiology, Netherlands Cancer Institute (NKI), \\1066 CX, Amsterdam, The Netherlands \and
GROW School for Oncology and Development Biology, Maastricht University, \\6200 MD, Maastricht, The Netherlands \and
Faculty of Applied Sciences, Macao Polytechnic University, 999078, Macao, China\and
Department of Radiation Oncology, Netherlands Cancer Institute (NKI), \\1066 CX, Amsterdam, The Netherlands \and
Department of Radiology and Nuclear Medicine, Radboud University \\Medical Centre, 6525 GA, Nijmegen, The Netherlands \and
Department of Biomedical Sciences and Morphologic and Functional Imaging, \\AOU G. Martino, University of Messina, 98100, Messina, Italy
\\
* Corresponding author: \email{taotanjs@gmail.com}}
\maketitle              
\begin{abstract}
Precision breast cancer (BC) risk assessment is crucial for developing individualized screening and prevention. Despite the promising potential of recent mammogram (MG) based deep learning models in predicting BC risk, they mostly overlook the “time-to-future-event" ordering among patients and exhibit limited explorations into how they track history changes in breast tissue, thereby limiting their clinical application. In this work, we propose a novel method, named \M, to precisely model the ordinal relationship of the time to and between BC events while incorporating longitudinal breast tissue changes in a more explainable manner. We validate our method on public EMBED and inhouse datasets, comparing with existing BC risk prediction and time prediction methods. Our ordinal learning method \M~outperforms existing methods in both BC risk and time-to-future-event prediction tasks. Additionally, ordinal heatmap visualizations show the model's attention over time. Our findings underscore the importance of interpretable and precise risk assessment for enhancing BC screening and prevention efforts. The code will be accessible to the public.


\end{abstract}
\section{Introduction}
Breast cancer (BC) screening, aiming to detect tumors at an early stage, can efficiently reduce the mortality of BC in women \cite{wang2022artificial}. However, the one-size-fits-all BC screening policy generates millions of negative mammograms (MG) annually worldwide and requires substantial resources \cite{yala2021toward}. To optimize BC screening, risk prediction models have emerged as promising tools for stratifying the screening population by guiding high-risk patients toward supplemental imaging or more frequent screenings while recommending longer intervals for low-risk populations to mitigate overdiagnosis \cite{wang2022artificial}.

Recently, research has explored artificial intelligence (AI) based risk models to leverage risk-related predictive patterns from MG \cite{wang2022artificial}, which achieved superior BC risk stratification compared to only risk factors based traditional risk methods \cite{eriksson2023long,gastounioti2022artificial,yala2023rethinking}. These methods focus primarily on identifying high-risk patients, where they deal with BC risk prediction as binary classification \cite{dadsetan2022deep,liu2020decoupling,yeoh2023radifusion} or as multi-binary classification tasks \cite{lee2023enhancing,wang2023predicting,yala2021toward}. Nevertheless, for practical medical applications, precise time-to-future BC prediction is more helpful for physicians in deciding prevention strategies or the timing of subsequent screenings. The existing models, however, cannot accurately predict when breast cancer occurs due to their lack of focus on the time-to-event estimation \cite{xiao2020censoring}. The time-to-event estimation task could guide the model to learn the natural ordering of the time to BC among the patients. Therefore, ideally, integrating both risk prediction and time estimation tasks enables not only the precision of time predictions but also augments the model's capability to identify features indicative of BC development more effectively. 

Furthermore, incorporating the historical images is vital for comprehensive learning of both the BC risk-related predictive patterns and time-to-cancer-related ordinal biomarkers. Several efforts have already demonstrated that multi-time-point (MTP) based methods \cite{wang2023predicting,yeoh2023radifusion,lee2023enhancing,dadsetan2022deep} can enhance the accuracy of BC risk assessment compared to single-time-point (STP) based approaches. Despite the advancement, most of them \cite{lee2023enhancing,dadsetan2022deep} leverage implicit (unsupervised) multi-view attention mechanisms (e.g., transformer-based approaches)
to infer information from previous scans indirectly with limited exploration of the explainable spatial-temporal analysis process. 
Furthermore, these methods may potentially introduce biases unrelated to the actual disease progress observed in multi-time-point images, such as variations in breast compression during imaging sessions.
Hence, there still exists a significant need for the explicit monitoring and tracking of temporal variations related to BC risk, which increases not only the robustness of the risk prediction models but also their explainability.

In this paper, we address the limitations described above by introducing a novel ordinal-learning-based risk model that uses longitudinal attention alignment.
Our contributions are summarized as follows:
\textbf{\textit{(1)}} We propose an ordinal learning framework for concurrently considering both time-to-event BC prediction and risk stratification tasks.
\textbf{\textit{(2)}} We incorporate attention alignment mechanisms to explicitly capture risk-related changes from multi-time point MG in an interpretable manner.
\textbf{\textit{(3)}} We validate the efficacy and robustness of our approach on two longitudinal MG datasets for future BC risk prediction.
Using the ordinal learning and interpretable attention mechanisms, our model offers a promising avenue for improving the accuracy and interpretability of BC risk prediction--both vital for personalized screening.

\section{Methodology}

\begin{figure}[!t]
\centering
\includegraphics[width=\textwidth]{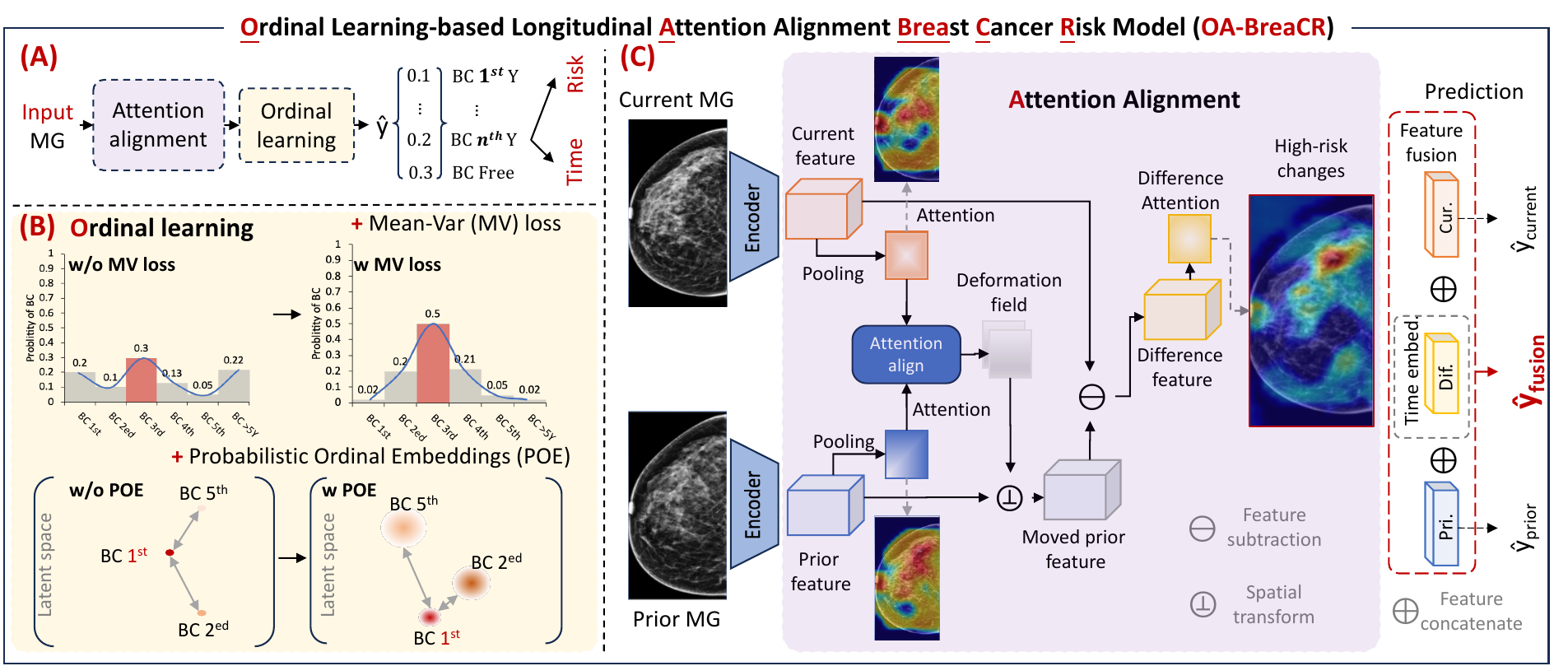}
\caption{The schematic overview of the proposed \M.
(A) Our method utilizes ordinal learning and longitudinal attention alignment for the risk and time-to-BC prediction tasks. (B) Ordinal learning for estimating the time to future BC events by leveraging Mean-variance (MV) Loss \cite{pan2018mean} and probabilistic ordinal embedding (POE) \cite{li2021learning}. (C) The attention alignment model aims to learn temporal breast tissue changes from two-time point mammograms in an interpretable manner.
} \label{fig:network}
\end{figure}

\subsection{Definition of the risk and time-to-BC prediction tasks}
\textbf{Risk prediction} This task aims to estimate the probability of future BC within a span of $n$ years subsequent to a specified MG, $\mathbf{x} \in \mathbb{R}^{H \times W}$, with $n=5$ in this study. As depicted in Fig \ref{fig:network} A, our proposed OA-BCR model, is designed to predict an $(n+1)$-dimensional vector $\hat{\mathbf{y}}$ of mutually exclusive outcomes. 
The initial $n$ outcomes indicate the probability of BC emergence in each subsequent year up to the $n^{\text{th}}$ year, while the final outcome denotes the probability of remaining cancer-free over the $n$-year timeframe.  
The BC risk within $m$ years ($m \leq n$),  denoted as $\text{Risk}_m$, is defined by $\text{Risk}_m = \sum_{i=1}^m \hat{\mathbf{y}}_i$, where $\hat{\mathbf{y}}_c$ is the probability of BC occurrence in the $i^{\text{th}}$ year. This approach builds upon methods found in prior research \cite{wang2023predicting}.

\textbf{Time-to-BC prediction} \textit{“If we identify the high-risk population, is it possible to estimate the onset time to future BC?"} 
Inspired by classification-based regression for predicting age \cite{pan2018mean}, we formulate time-to-event prediction as an ordinal regression problem. This is defined by, $\hat{\mathbf{t}}=\sum_{i=1}^{n+1} i \; \hat{\mathbf{y}}_i$, where $\hat{\mathbf{t}}$ is estimated time to BC onset as determined by the risk model. To enhance the accuracy of the time prediction, we have integrated two regression-task-specific functions aimed at constraining the risk model to learn the ordinal dynamics inherent in the progression toward BC.

For each sample, the corresponding ground truth, $\mathbf{y}$, is an $(n+1)$-dimensional binary vector.
The binary cross-entropy (BCE) loss for the $i^{\text{th}}$ category of each sample is calculated as $\ell^{\text{bce}}(i) = -\mathbf{y}(i) \log(\hat{\mathbf{y}}(i)) - (1-\mathbf{y}(i)) \log(1-\hat{\mathbf{y}}(i))$.
To accommodate samples with incomplete follow-up data, a masking function, $\delta(i)$, is employed to determine the inclusion of a sample at class $i$ in the loss computation. It is defined as \textit{$\delta(i) = 1$, if the follow-up period $\geq$ min(i, n) or diagnosed within $n$ years, and $0$ otherwise}.
This method allows for the exclusion of periods with unknown outcomes, concentrating solely on accessible data. Consequently, the overall BCE loss, $L_{\text{BCE}}$, is computed as $L_{\text{BCE}} = \sum_{i=1}^{n+1} \delta(i) \ell^{\text{bce}}(i)$, ensuring the loss accurately reflects the available follow-up label for each sample.

\subsection{Ordinal learning for estimating the time to future BC events}
\textbf{Mean-Variance (MV) Loss} Motivated by the existing ambiguity of the tumor stage when patients are diagnosed, the probability of getting a BC should increase as the diagnostic date nears. Therefore, the predictive model should learn the distribution of cancer occurrence times, aiming for the estimated probability distribution to be close to a normal distribution. As shown in Fig \ref{fig:network} B, the maximally probable time should approximate the actual time ($\mathbf{t}$) of diagnosis; the predicted probabilities should diminish as it moves away from $\mathbf{t}$. 
Thus, we employ the MV loss function \cite{pan2018mean} to constrain the mean and variance of the predicted probability distribution.
The MV loss is presented as $\mathcal{L}_{\text{MV}} = \Vert \hat{\mathbf{t}}-\mathbf{t} \Vert_2^2 + \ell^{\text{var}}(\hat{\mathbf{y}})$, where $\ell^{\text{var}}$ is variance loss introduced by \cite{pan2018mean}.
It aims to ensure that the estimated time aligns closely with the actual time of diagnosis constrains the probability distribution as concentrated as possible.

\textbf{Probabilistic ordinal embedding (POE)} In addition to constraining the predictive distributions, we also integrate the POE technique \cite{li2021learning} to reinforce the ordinal relationships learning within the latent space. As shown in the bottom of Fig. \ref{fig:network} B, this method represents each sample as a multivariate Gaussian distribution rather than a deterministic point in the latent space. Furthermore, it takes into account the ordinal relationships among MGs of women diagnosed with BC at varying years post-scan (e.g., first, second, and fifth year), learning a distribution following the ordinal constraint (e.g., $1 < 2 < 5$). The POE loss can be defined as $\mathcal{L}_{\text{POE}} = \ell^{\text{ord}} + \ell^{\text{KL}}$. The triplet loss ($\ell^{\text{ord}}$) \cite{li2021learning} enforces the ordinal relationships within the embedding space, ensuring that the learned representations maintain the correct sequence order. The KL divergence term ($\ell^{\text{KL}}$) is used to constrain probabilistic ordinal embeddings for minimizing the divergence between the predicted and standard normal distributions. 

\subsection{Risk-related attention alignment of prior-current mammograms}
To learn both the BC risk-related predictive \cite{lee2023enhancing,wang2023predicting,yala2023rethinking} and time-to-cancer-related ordinal patterns, the model must capture features related to breast tissue change from multi-time-point images.
However, the inherent two-dimensional projection principle of mammography introduces challenges in spatial-temporal analysis due to significant deformations between two time points arising from variations in breast compression during imaging sessions. Image-level registration methods may struggle to handle such large deformations accurately, and incorrect registration may mislead the prediction model \cite{han2024deform,dong2023preserving}. To address this challenge, we propose the attention alignment module based on tissue features.

\textbf{Attention alignment}
Unlike transformer-based methods which implicitly fuse the prior and current features \cite{lee2023enhancing,wang2023predicting}, our method aims to explicitly compare the two images in an explainable way, as shown in Fig \ref{fig:network} C. Utilizing an attention-guided pooling method \cite{mikhael2023sybil}, we obtain attention maps $\{a^{\text{cur}}, a^{\text{pri}}\}$ 
corresponding to the $\{f^{\text{cur}}, f^{\text{pri}}\}$ for which both $f \in \mathbb{F}^{h \times w \times c}$
derived from the encoder. The attention maps indicate the high-risk areas of the corresponding images. Subsequently, these paired prior-current attention maps are fused within an attention alignment block, which is a convolutional block (two convolutional layers with batch normalization and ReLU activation functions), designed to estimate the deformation field ($\phi$) of them. Inspired by Voxelmorph \cite{balakrishnan2019voxelmorph}, the alignment process involves registering the prior (moving) feature $f^{\text{pri}}$ to the current (fixed) feature $f^{\text{cur}}$ via a spatial transformation function \cite{balakrishnan2019voxelmorph}, resulting in the moved feature $\hat{f}^{\text{pri}}$. Here, we leverage the $L^2$ loss to constrain the registration processing, denoted as $\mathcal{L}_{\text{reg}} = \Vert \hat{f}^{\text{pri}},f^{\text{cur}}) \Vert_2^2$. Inspired by \cite{dadsetan2022deep,loizidou2021digital}, our method calculates the differential feature via subtraction of the current feature $f^{\text{cur}}$ and deformed prior feature $\hat{f}^{\text{pri}}$, which already aligned to $f^{\text{cur}}$. The difference attention map $a^{\text{dif}}$ can straightforwardly derived through the attention-guided pooling. 

\textbf{Multi-level (ML) joint learning}
To ensure the consistency of attention maps across images taken at different time points, we employ a multi-level (ML) joint learning strategy.
Specifically, current and prior features ($f^{\text{cur}}, f^{\text{pri}}$) are also used to predict the risks associated with each individual time point, followed by the computation of loss based on the labels corresponding to each image. By doing so, the attention mechanism could learn the progress knowledge regarding breast cancer (BC) development.
Subsequently, the final predictive modeling leverages the processed features derived from pooled vectors of $f^{\text{cur}}$, $\hat{f}^{\text{pri}}$ and $f^{\text{dif}}$. Note that, before feature fusion, time gap information between images is integrated into the differential vectors using position embedding techniques \cite{wang2023predicting}, enhancing the model's ability to account for temporal dynamics in risk assessment. This approach facilitates the comparison analysis of two-time point MG, thereby enabling an interpretable examination of temporal breast tissue changes.

\section{Experimental}
\subsection{Experimental settings}
\textbf{Datasets}
In this study, two longitudinal datasets were collected.
Firstly, the EMBED dataset \cite{jeong2023emory} is a large public screening dataset collected from two cohorts over an 8-year period.
We included all images with enough five-year follow-up results according to the released data, including 53,324 images from 6,571 patients.  
We partitioned the dataset by the patient to create training, validation, and test sets with a ratio of 5:2:3. The test set contains 15,720 images, including 574 cancer diagnosed within 5 years. 
Secondly, the Inhouse dataset comprises 171,168 images from 9,133 women between 2004-2020, collected from a hospital with IRB approvals. The dataset is randomly split into training, validation, and testing sets at the patient level in an 8:1:1 ratio. The test dataset includes 25,244 images consisting of 1,706 images diagnosed with cancer in five years. Details of time-to-cancer label distribution and dataset split are provided in the supplementary.

\textbf{Evaluation}
Following previous studies \cite{lee2023enhancing,yala2021toward}, we utilized C-index \cite{uno2011c} and the Area Under the Curve (AUC) to evaluate the performances of the methods in the 1-5 year risk prediction task. To evaluate the performance of time-to-event prediction, we used the mean absolute error (MAE) and cumulative score (CS) \cite{li2021learning,pan2018mean}. The MAE quantifies the average deviation between predicted and actual event times. The CS evaluates the precision within a specified tolerance (e.g., 1 year) of absolute error.
Due to the imbalance of the label distribution, we calculated both class-weighted MAE and CS, referred to as WMAE and WCS. Statistical significance is assessed through bootstrapping, with 1000 iterations, to establish confidence intervals as mean ± 1.96 standard deviations \cite{lee2023enhancing}.

\begin{table}[!b]
    \centering
    \caption{Comparisons with other methods in time-to-event estimation and risk prediction tasks. The methods with \textcolor[RGB]{234,180,
    138}{\textbf{orange color}} represent the single time point models, while \textcolor[RGB]{148,169,216}{\strut \textbf{blue color}} refers to the multi-time point models. The ± refers to the 95\% CI.}
    \includegraphics[width=.95\textwidth]{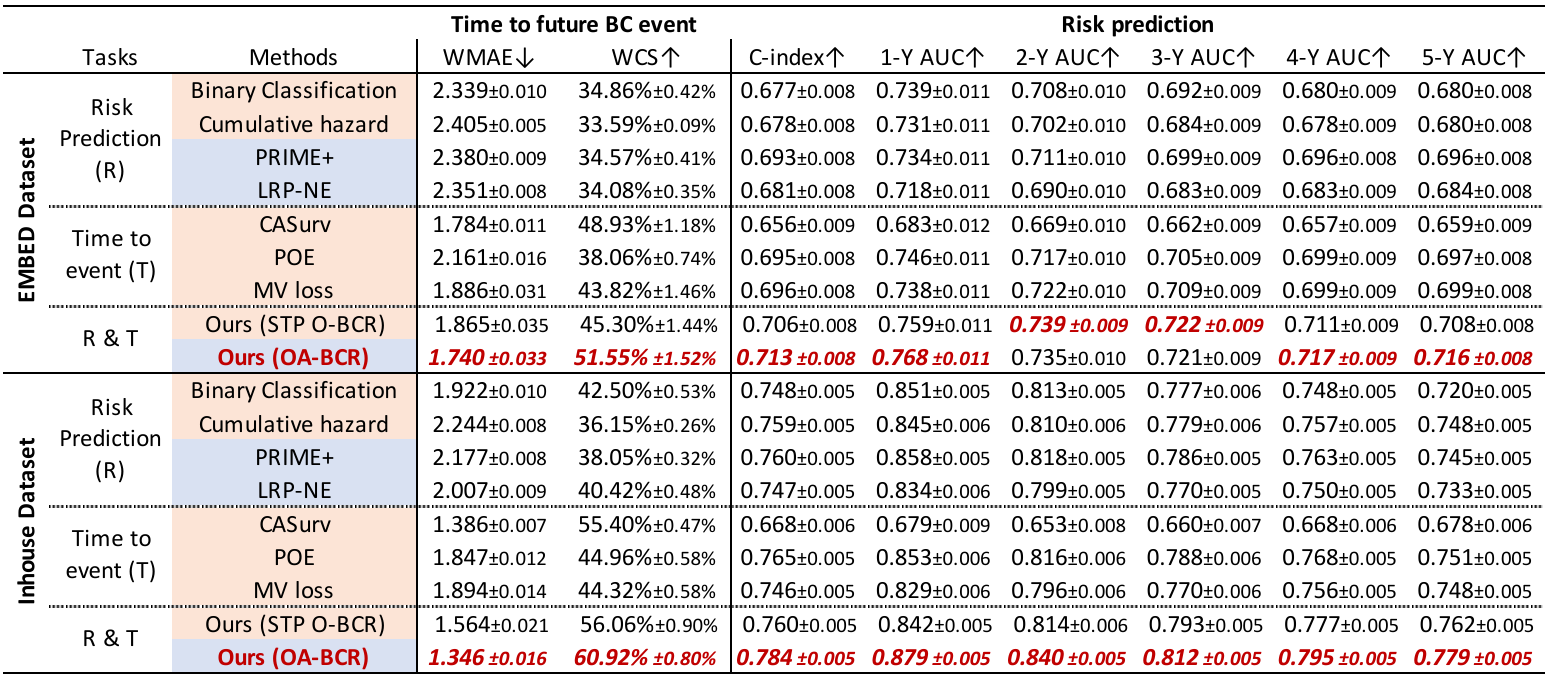}
    \label{tab:Comparison}
\end{table}

\begin{figure}[!b]
\centering
\includegraphics[width=.9\textwidth]{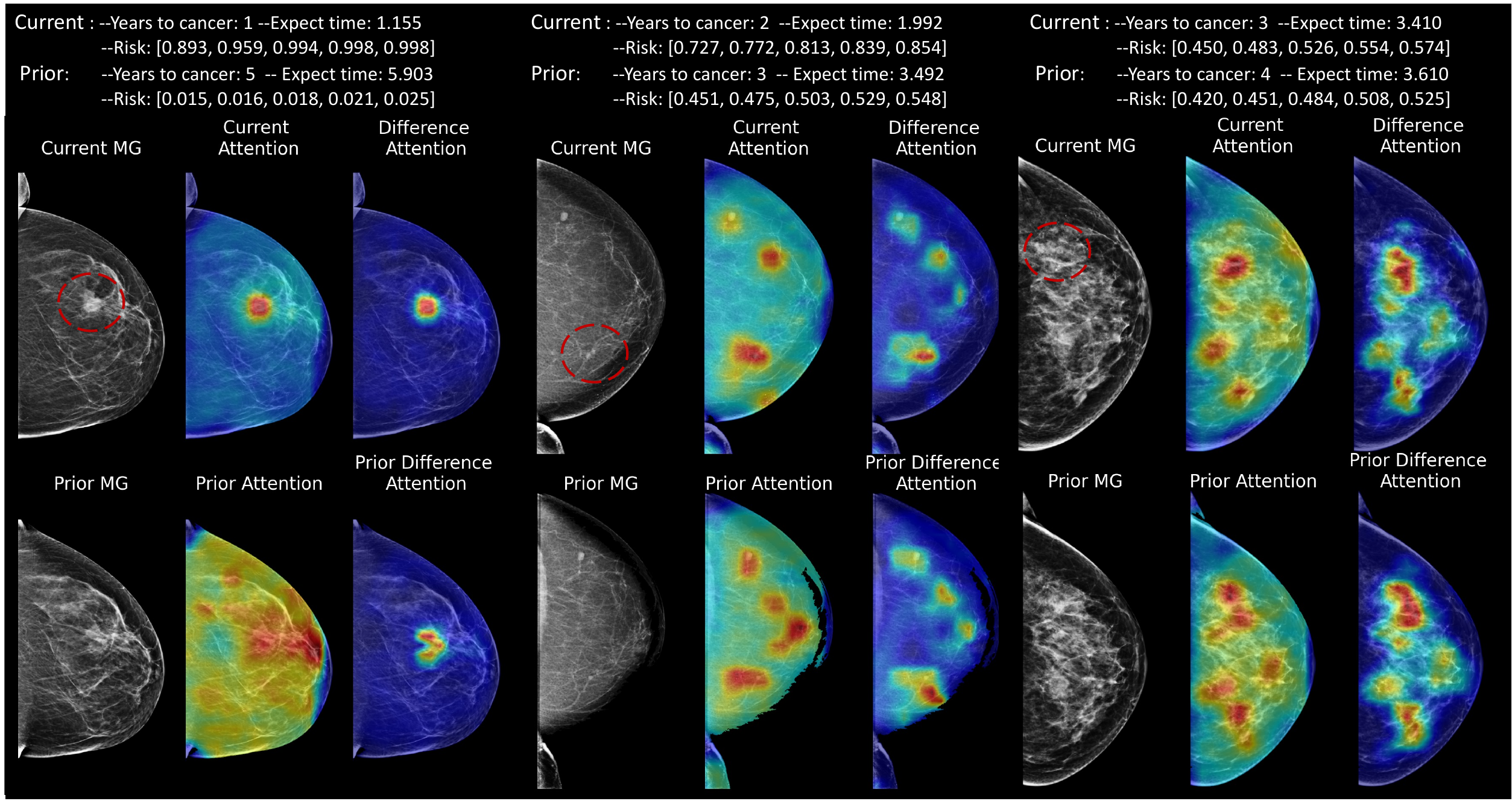}
\caption{Heatmaps of attention alignment on prior-current MG. The \textcolor{red}{\textbf{red}} circles indicate the the existing tumors or high-risk areas that developed the cancer within five years.} \label{fig:vis}
\end{figure}

\textbf{Implementation details}
All mammograms are preprocessed following the pipeline proposed by \cite{liu2020decoupling,wang2023predicting}, including segmenting the breast area, and resizing to 1024$\times$512 pixels of the image size while maintaining aspect ratio. 
To avoid overfitting, we leveraged augmentations including random flipping, rotation, and affine transformations (e.g., translations, and scalings) for training. We adopted ResNet-18 \cite{he2016deep}, initialized with ImageNet pre-trained weights, as the backbone of our model. During the training, the Adam optimizer is used, with a learning rate of $1\times10^{-4}$ and a batch size of 96. The learning rate is decayed with a rate of 0.5 and the patience is 5 continuous epochs that the metrics do not improve on the validation set. A total of 200 epochs for training and early stopping based on the validation c-index is utilized with patience of 15 continuous epochs. All deep learning models are implemented in the Pytorch framework and an NVIDIA RTX A6000 GPU (48GB), which takes 16-32 hours for training for each model.

\subsection{Experimental results}

\begin{table}[!t]
    \centering
    \caption{Ablation results in time-to-event estimation and risk prediction tasks. The methods with \textcolor[RGB]{234,180,
    138}{\textbf{orange color}} represent the STP models, while \textcolor[RGB]{148,169,216}{\strut \textbf{blue color}} refers to the MTP-based methods. The ± refers to the 95\% CI. MV, Mean-variance loss; POE, Probabilistic ordinal embedding; A2, Attention alignment; and ML, Multi-level.}
    \includegraphics[width=.95\textwidth]{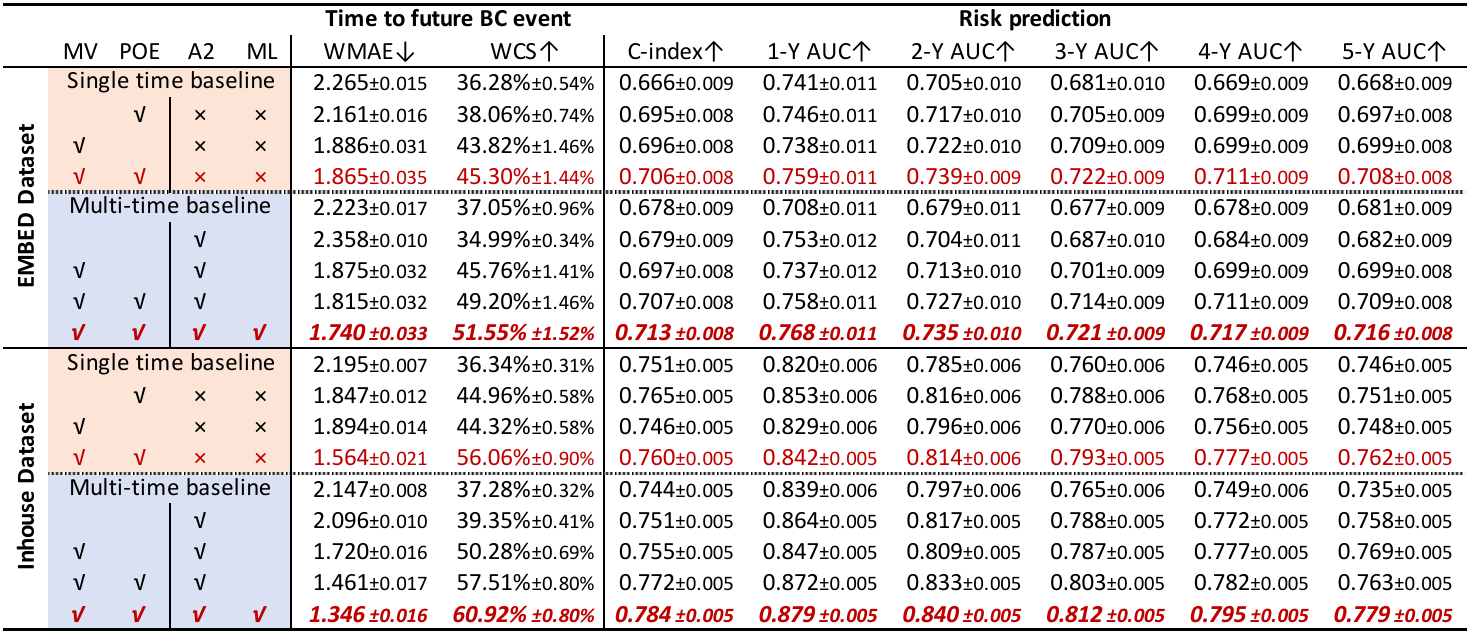}
    \label{tab:Ablat}
\end{table}

\begin{figure}[!t]
\centering
\includegraphics[width=.95\textwidth]{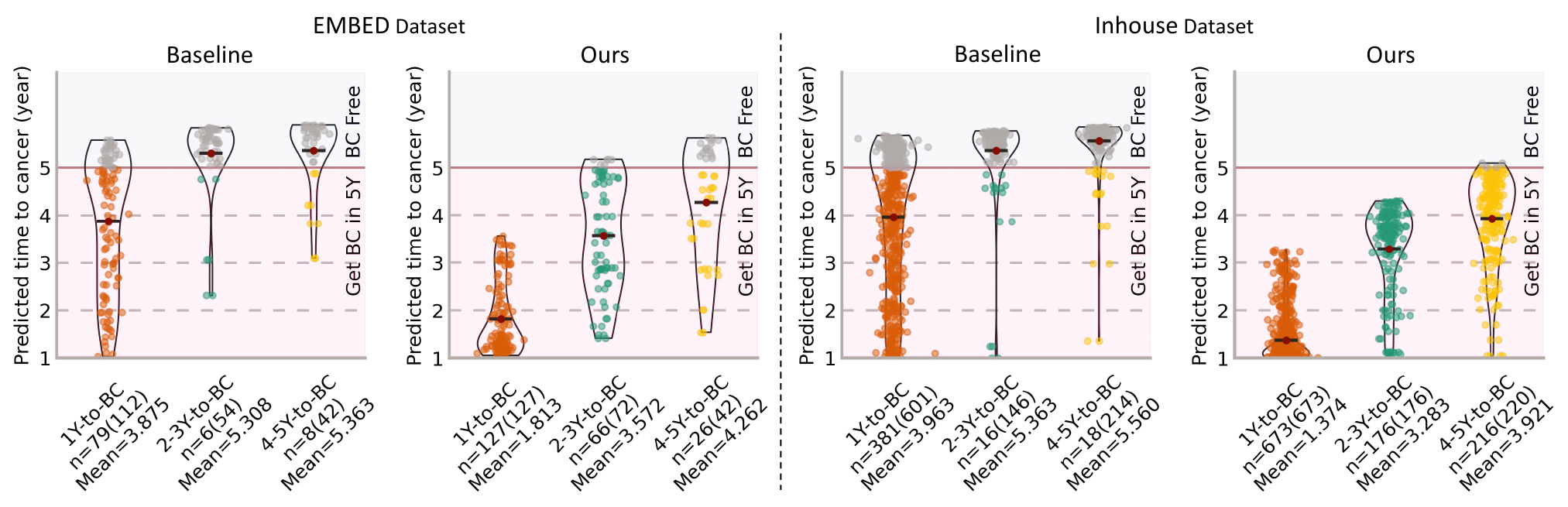}
\caption{Violin plots show the distributions of the expected time to cancer in stratified three high-risk populations. All dots indicate true high-risk patients whom the risk model successfully identified. Colored dots represent patients identified by the risk model with an estimated time to cancer within 5 years. While grey dots represent cases where the estimated time-to-cancer exceeds five years, indicating failures to accurately estimate time-to-cancer. \textbf{n} is the number of color dots, with the total count of color and grey dots provided in parentheses.
} \label{fig:dist}
\end{figure}

\textbf{Comparison of risk and time-to-cancer prediction.}
The comparisons are shown in Table \ref{tab:Comparison}, including the comparison of both STP and MTP methods. 
Within the STP framework, we compare our method with existing BC risk prediction methods, including binary classification \cite{liu2020decoupling}, and cumulative hazard method proposed by \cite{yala2021toward}. The time prediction models are compared with CASurv \cite{hermoza2022censor}, POE \cite{li2021learning}, and MV \cite{pan2018mean} methods.
For the MTP-based method, we compare our approach with a transformer-based approach (PRIME+) \cite{lee2023enhancing} and a pre-registration based RNN based method (LRP-NE) \cite{dadsetan2022deep}. 
The results show for the STP setup, our proposed ordinal learning yields an observable improvement on the time prediction with a stable advantage for the risk prediction. More importantly, our proposed method, \M, demonstrates a better performance across both the risk and time prediction tasks, as well as across both evaluated datasets, compared to all other considered methods. 

\textbf{Visualization of attention alignment.}
Fig. \ref{fig:vis} displays multiple attention-aligned cases. 
The attention of the baseline method is globally on the dense breast tissue, consistent with findings of previous research \cite{liu2020decoupling}.
However, after the attention alignment module, even without pre-registration, the heatmaps show that the attention of the model accurately focuses on the local changes in the breast tissue.

\textbf{Ablation study.}
To better understand the effectiveness of the proposed module in this study, we also conduct an ablation study, as shown in Table \ref{tab:Ablat}. For the SPT-based method, we ablate utilized MV and POE of the proposed ordinal learning method. 
For the MTP-based method, we ablate both the ordinal learning part and the attention alignment module. The attention alignment module includes attention alignment (A2) and multi-level (ML) learning. The results show that our proposed ordinal learning model improves the precision of the time prediction. The advantage is further improved when cooperating with the prior image through our proposed attention alignment method. The conclusion is consistent on both the public and inhouse datasets.

\textbf{Comparison of time-to-future-BC estimation in the detected true high-risk groups.} To further investigate the abilities of the time to cancer estimation, we plot the distribution of the expected time for the detected high-risk population. Fig \ref{fig:dist} shows that our proposed ordinal learning not only could detect more high-risk patients but also could improve the model's precision for time-to-cancer estimates on both two datasets.

\section{Conclusion}
In this study, we introduce an ordinal learning for explainable multi-time-point BC risk prediction model, \M. Our method incorporates both prior mammograms and current mammograms with an attention alignment module to explicitly capture changes in breast tissue over time. More importantly, ordinal learning could further improve the precision of risk prediction, enhancing the ability to predict time-to-events. The experiments on two large datasets demonstrate the robustness of our method. Our findings highlight the importance of interpretable and precise risk assessment for enhancing BC screening and prevention efforts. The accuracy of the high-risk area detection can be further validated, which may improve the credibility of the risk model in the clinical practice of BC screening.

\begin{credits}
\subsubsection{\ackname}
Xin Wang is funded by Chinese Scholarship Council scholarship (CSC) and this work was also funded by the Science and Technology Development Fund of Macau SAR (Grant number 0105/2022/A). The authors would like to acknowledge the Research High Performance Computing (RHPC) facility of the Netherlands Cancer Institute (NKI).

\subsubsection{\discintname}
The authors declare no competing interests.
\end{credits}
%
%
%
\bibliographystyle{splncs04}
\bibliography{ref}
\end{document}